\documentclass[prb,twocolumn,showpacs]{revtex4}

\usepackage{graphicx}

\begin{document}

\title{Characterisation of high energy electron irradiation damage in UPt$%
_{3}$ samples}
\author{P.\ Rodi\`{e}re$^{\dag \ddag}$, J.P. Brison$^{\sharp}$, A.D.\ Huxley
$^{\dag}$, F. Rullier Albenque $^{\natural}$, and \ J.\ Flouquet $^{\dag}$}
\address{$\dag$ D\'{e}partement de Recherche Fondamentale sur la Mati\`{e}re Condens%
\'{e}e, SPSMS, CEA-CENG, 38054 Grenoble-Cedex 9, France\\
$\ddag$ H.H. Wills Laboratory of Physics University of Bristol Tyndall
Avenue Bristol, BS8 1TL U.K.\\
$\sharp$ CRTBT, CNRS, 25 avenue des Martyrs, BP166, 38042 Grenoble Cedex 9,
France \\
$\natural$ SPEC, Ormes des Merisiers, CEA, 91191 Gif sur Yvettes, France}

\begin{abstract}
We present transport and specific heat measurements on high
quality  single crystals of UPt$_3$ before and after irradiation
by high  energy electrons. We observe a strong dependence of the
critical  temperature with the sample thickness. The dramatic
effects of the  irradiation on the specific heat in the
superconducting state can  then be simply explained in terms of an
inhomogeneous distribution of superconducting transition
temperatures. The question of the failure of the "universal limit"
in the heat transport of UPt$_3$ is reexamined, and the conditions
for a clean experimental test are established.
\end{abstract}

\maketitle




%


\section{Introduction}

Unconventional superconductors are mainly found among strongly
correlated electronic systems,  where impurities seem to be
dominantly in the unitary limit \cite{Pethick86}. In such a case,
even a very small concentration of
defects will induce a finite density of states at zero energy \cite%
{Scmitt-86}. Moreover, a finite impurity concentration induces conduction
bands which contribute to transport properties \cite{Balatsky95}: for
example, impurities should  add a linear term in temperature (T) to the
thermal conductivity for T$\rightarrow 0$. More  subtle effects were
predicted depending on the particular gap topology. If a line of nodes is
present on the Fermi surface, and in presence of defects, the  linear term
of the thermal conductivity should be universal ( i.e. independent of the
impurity  concentration) for a small variation of T$_{c}$ \cite%
{Lee-93,Graf-99}. This is considered as a  very robust test of the
unconventional nature of the superconducting state and of its gap topology.

Such a universal limit has been observed in a few quasi-2D unconventional
superconductors such as  YBa$_2$Cu$_3$O$_{6.9}$ \cite{Taillefer-97}, Bi$_2$Sr%
$_2$CaCu$_2$O$_8$ \cite{Behnia-99} and  Sr$_2$RuO$_4$ \cite{Suzuki-02}. But
it failed in the underdoped High-T$_c$ compound YBa$_2$Cu$_4$O$_8$ \cite%
{Hussey-00}. In this last case, the origin of the absence of the universal
limit is still  not clear, and could be attributed among others to the
nature of the superconductivity or the  particular ground state of this
compound. Further investigations have to be done in compounds where the
electronic ground state is well established. Moreover, the universal limit
has never been  observed in any three dimensional unconventional
superconductor.

One of the best studied unconventional 3D superconductor is the heavy fermion UPt$_3$. Its  superconducting state is
particularly interesting because it exhibits a well known phase diagram \cite{Hasselbach-89,Adenwalla-90}, each phase
corresponding to a different symmetry of the  superconducting gap. Various scenarios exist for these phases
\cite{Joynt-02},  corresponding to different types of nodes of the superconducting gap. Probing the universal  limit in
the different
crystallographic directions would be a very strong test of the models \cite%
{Graf-99}.

Impurity effects in UPt$_3$ have been studied in many different
ways. Substitution of Uranium or Platinum by another element have
been used
to study the decrease of the critical  temperature \cite%
{Vorenkamp-93,Dalichaouch-95,expGraf-99}, and the interplay between the
magnetism and the superconductivity \cite{deVisser-02}. They would not be
appropriate to probe  the universal limit due to the rapid change of the
normal state properties and T$_{c}$ with even  the smallest concentrations
that can be reasonably studied. Small defect concentrations can be  tuned
with annealing, which varies the number of intrinsic metallurgical defects.
A main  advantage of annealing is the possibility to follow the effects on
the same sample. Its  influence on physical properties at low temperature
like the specific heat \cite{Brison-94} or  transport properties \cite%
{Kycia-98} has been reported. Nevertheless, it does not seem  appropriate to
investigate a possible universal limit in UPt$_3$ due to the small range of
change  of the critical temperature and the lack of knowledge of the nature
of the defects : extended  defects are expected rather than point defects,
and they may require more involved descriptions \cite{Kycia-98} which spoil
the strength of the theoretical predictions.

A third technique uses high energy electron irradiation damage to
create point defects \cite{Suderow-99}. It seems ideal for such
studies, combining the advantages of the previous techniques : the
vacancy and interstitial defects created by electron irradiation
have a size of a few \AA ngstr\"{o}ms, small compared to the
coherence length of UPt$_{3}$ ($\xi _{0}\approx 100$\AA );
Moreover, it is possible to access a wide range of defect
concentrations simply by tuning the irradiation time. Finally, the
same sample can be studied for the various concentrations, and a
simple annealing at 700$^{\circ }$C suppresses again all defects,
refreshing the sample. This technique has been used in UPt$_{3}$
to study the influence of impurities on T$_{c}$, on the upper
critical field H$_{c2}$ and on the thermal conductivity at very
low temperature \cite{Suderow-99}. No ''universal limit'' has been
observed, a surprising result given the fact that the thermal
conductivity of the pure sample is perfectly understood with the
most popular scenarios \cite{Lussier97,Suderow-97} which predict a
universal behaviour at least in the basal plane. But measurements
of the specific heat of irradiated samples already indicated a
very unexpected behaviour, with a large broadening of the
transition \cite{Brison00}. Theoretical investigations of the
specific heat broadening due to the natural (gaussian) dispersion
of defect concentrations \cite{Fomin00} cannot explain the
importance of the effect observed, and further experimental
characterisation of the induced disorder seemed necessary.

In this paper, we present new results on the specific heat and
transport measurements of UPt$_3$ samples damaged by high energy
electrons. We show that the sample thickness has a dramatic
influence on the results, explaining the previous failure of the
measurements \cite{Suderow-99}. We give quantitative estimates of
the distribution of defects in the sample, and determine the
conditions for a clean test of the universal limit in this system.
The experimental procedure and the main results are described in
section II. In section III, an introduction to the creation of
defects by electron irradiation is presented. Section IV presents
a phenomenological model used to describe quantitatively our
results. In the last sections, the previous results are briefly
discussed in the framework of our description.\newline

\section{Experimental details and main results}

Pure samples were cut from the same single crystal grown under
ultra-high vacuum by the Czochralski method. The shape of the
sample for resistivity measurements was a bar of typical length
5~mm and width 0.23~mm. The samples R1, R2 and R3 have been
thinned down to 190, 90 and 70 $\mu$m respectively, in order to
probe the depth of the irradiation damage. R4 and R5 with a
respective thickness of 220 $\mu$m and 240 $\mu$m were prepared to
study the evolution of the normal state transport properties for
different rates of irradiation. R1 and R4 were cut with the
electrical current along the c-axis, R2, R3 and R5 with the
electrical current in the basal plane. The two samples measured by
specific heat (C1 and C2) were plates 270~$\mu$m thick and a
surface roughly 2x2~mm$^2$. After being cut and/or thinned, all
samples were annealed under ultra high vacuum ( $< 3.10^{-10}$ Torr) at 950$%
^{\circ}$C during 5 days. The quality of the resistivity samples has not been altered by the thinning process since the
three samples exhibited a high critical temperature $T_c$ of 0.55~K (R1) and 0.56~K (R2 and R3).

The irradiations by high energy electrons have been performed in
the low temperature facility of the Van der Graaff accelerator of
the ''Laboratoire des solides irradi\'{e}s'' of Ecole
Polytechnique at Palaiseau (France). The resistivity samples were
held next to each other on a thin fibre glass plate with GE
varnish. The thinnest dimension was parallel to the electron beam.
This plate was wrapped in a thin copper foil, held rigidly inside
a liquid hydrogen bath. The three samples were irradiated
simultaneously, and they remained well within the homogeneous
region of the electron beam. Electrical contacts on the samples
were done with silver paste (solder might have reannealed part of
the defects), removed before, and made again after the
irradiation. The samples C1 and C2 for the specific heat
measurements were wrapped directly the thin copper foil, and were
irradiated separately. Incident electrons were accelerated at an
energy of 2.5~MeV for samples for resistivity measurements and C1,
and with an energy of 1.5~MeV in the case of sample C2. The
irradiation takes place in liquid hydrogen. Two Faraday cups
located in the front and behind the samples are used to estimate
the irradiation fluence at the level of the sample
\cite{Gosset-87}. The electron flux is limited to
$\sim$10$^{14}$~e$^{-}$/cm$^{2}$/s in order to avoid the heating
of the sample during irradiation. Fluences ($\Phi $) between $1.8$
and $13.9\times 10^{18}$ electron/cm$^{2}$ (e$^{-}$/cm$^{2}$) have
been used. After irradiation, the samples were warmed up to room
temperature and transferred in different experimental set-ups to
perform resistivity and specific heat measurements. This procedure
results in some annealing of the defects created at low
temperature.

%
%

\begin{figure}[t]
\includegraphics[bb=54 241 530 598, width = 0.9\linewidth,
clip]{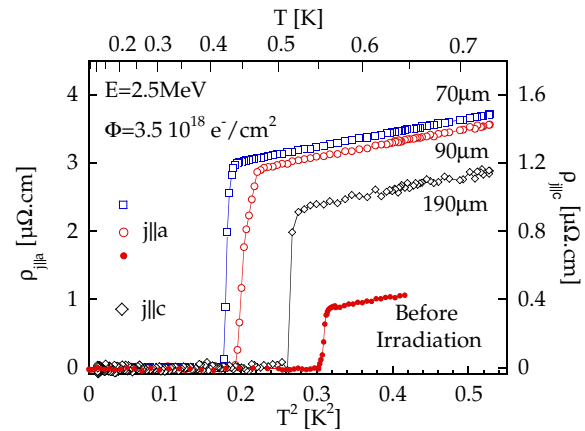} 
\caption{ Temperature dependence of the resistivity of the three
samples of
UPt$_3$, R1 R2 and R3 (with a thickness of respectively 190, 90 and 70 $%
\protect\mu$m) after an irradiation by electrons of energy of 2.5
MeV with a fluence of 3.5 10$^{18}\times$e$^-$/cm$^2$. The sample
R2 is also shown
before irradiation (closed circles) and is typical of pure samples of UPt$_3$%
. The current is oriented along the a-axis except for sample R1
(c-axis). The left scale respects the ratio of the anisotropy of
the effective mass of the quasi-particules observed in pure
samples (2.5). } \label{figRes}
\end{figure}

The resistivity measurements have been performed with a traditional four
wires technique with a 17~Hz AC-current bridge. The specific heat
measurements have been performed with a semi adiabatic technique. All
samples have been measured before and after irradiation.

When electrons go through the matter, a part of their energy is lost via inelastic collisions with the target
electrons. Typically, the energy loss is of the order of 1-2 keV/$\mu $m \cite{Pages}. In the case of 100-200 $\mu $m
thick the energy loss can be as high as 400~keV. But as long as the energy transferred to the target nucleus, which
depends on the atomic weight of the nucleus and on the incident energy of the electron, is much larger than the minimum
energy required to displace it (threshold energy), the cross section for the atomic displacement does not depend
drastically on this energy. So, in this case, the decrease of the electron energy through the target compound will not
induce strong inhomogeneities in the damage distribution. However, if the energy transferred to the target nucleus is
close to the threshold energy, one could expect that the decrease of the incident energy along the path induces that
electrons will not be longer able to displace atoms in the whole thickness of the samples. In this latter case the
critical temperature can be expected to change differently along the electron path depth. We suspect that this
situation could explain the large broadening of the specific heat transition observed previously on irradiated
UPt$_{3}$ samples \cite{Brison00}. In order to test this assumption, we have performed two kinds of measurements : (i)
we have used lower energy electrons (if close to the threshold, the irradiation effects should then be drastically
reduced), and (ii) we have measured irradiated samples of various thicknesses in order to check a possible dependence
of T$_{c}$ with thickness.

Figure \ref{figRes} shows the result of the last test. It displays
the resistivity of the three samples of different thickness,
irradiated during the same experiment and with the same fluence
$\Phi =3.5\times 10^{18}$e$^{-} $/cm$^{2}$. The most striking
result is that indeed, the thinner the sample, the stronger the
drop in T$_{c}$. This correlation between sample thickness and
T$_{c}$ change is a direct proof of the inhomogeneity of the
irradiation damage. As regard the normal state properties, the
temperature dependence of the resistivity remains quadratic ($\rho
(T)=\rho _{0}+A.T^{2}$) for all samples. Both terms $\rho _{0}$
and $A$ increase under irradiation, and again, $\rho _{0}$
increases more strongly when the sample is thinner. However, $A$
is closer to the value of a pure sample when the irradiated sample
is thinner. We take this as pointing to an increased average
disorder for thinner samples and a wider distribution of defects
for thicker samples. We shall make these conclusions more
quantitative in section V.

%
%

\begin{figure}[t]
\includegraphics[bb= 71 107 502 718, width = 0.6\linewidth,
clip]{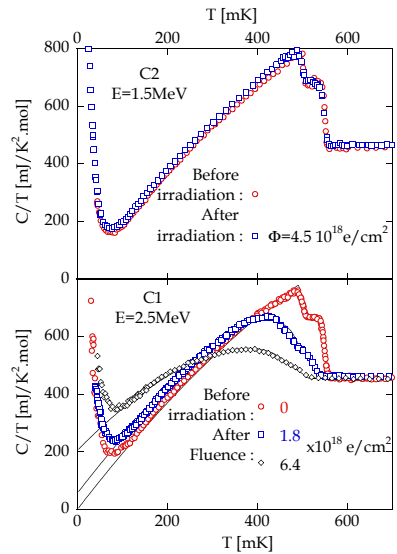} 
\caption{Temperature dependence of the specific heat of the same
sample of UPt$_3$ before (open circle) and after damaged by
irradiations. The irradiations were performed with incident
electrons at an energy of 1.5MeV (top panel sample C2) and 2.5 MeV
(bottom panel sample C1). The top panel demonstrate that at
1.5MeV, no effects are observed on the specific heat (no defects
creation). On the bottom panel, the line for the sample before
irradiation is a phenomenological description of the specific heat
behaviour with a small homogeneous distribution of $T_c$ of 16mK.
After irradiation, a good fit in the whole temperature range and
for the two fluence is obtained with an exponentional decrease of
the rate of defects in the depth of the sample (see text). The
exponential distribution is characterised by the length
$\protect\lambda=90~\protect\mu$m and the decrease of $T_c$ for an
infinitely thin sample by the parameter
dT$_c$/d$\Phi$=-115~mK/10$^{18}$ e$^- $/cm$^2$. } \label{figCp}
\end{figure}

Figure \ref{figCp} presents the effects of irradiation at two different
energies on the specific heat. The top panel compares virgin sample C2, and
the same sample irradiated with a fluence of $4.5\times10^{18}$e$^-$/cm$^2$
at 1.5~MeV : within experimental accuracy, there is no effect at all of the
irradiation, both in the normal and superconducting state. The simplest
interpretation is that with an incident energy of 1.5~MeV, the transmitted
energy to the uranium and platinum atoms is below the threshold energy for
defect formation. These results at 1.5~MeV should be contrasted with the
bottom panel of figure \ref{figCp}. It shows the drastic effects at 2.5~MeV %
\cite{Brison00} on sample C1, with a new measurement at a low fluence of $%
1.8\times10^{18}$e$^-$/cm$^2$. The virgin sample C1 (like C2) exhibits the
well known double specific heat jump at temperatures (defined by an onset
criteria) of T$_{c^+}$=550~mK and T$_{c^-}$=498~mK respectively. Both
transitions show only a small broadening of 16 and 12~mK respectively,
demonstrating the very good quality of the sample. Note that at very low
temperature the strong upturn in the specific heat already reported \cite%
{schuberth-94} is well observed. The origin of this upturn could be a
magnetic ordering transition but is not directly connected to the
superconducting state. In the following analysis this upturn will be
ignored. After irradiation, the same sample shows a strong broadening of the
specific heat : the two superconducting transitions are already hardly
distinguishable for a fluence of $1.8\times10^{18}$e$^-$/cm$^2$, and are
totally smeared at a higher fluence.

By contrast, the onset critical temperature shifts only very
little, much less than observed by resistivity on thin samples (R3
for example). From this point of view, the irradiation effects are
different from the annealing effects where broadening and T$_{c}$
decrease are intimately correlated \cite{Brison-94}. It points
again to an inhomogeneous distribution of point defects. Let us
note that normal state properties are little affected by
irradiation (apart from a smaller mean free path). Indeed, above
the superconducting transition, the sample exhibits a clear linear
dependence of the specific heat ($C=\gamma T$) with a large value
of the Sommerfeld coefficient $\gamma $ characteristic of the
heavy fermion Fermi liquid regime. The irradiations do not affect
this Fermi liquid behaviour nor the value of $\gamma $. They are
also reversible : the original specific heat has been recovered
after a subsequent annealing at 700$^{\circ }$C during 5 days
under ultra high vacuum.\newline

\section{Creation of defects by high energy electron irradiation}

We have stated that the experimental behaviours of the resistivity and the specific heat suggest that 2.5~MeV is very
close to the energy necessary to create stable defects in UPt$_{3}$. In this section, we will present a short
introduction to the creation of defects by high energy electron irradiation with a special emphasis to the special case
of UPt$_{3}$. A plausible qualitative explanation of the origin of such inhomogeneities is proposed. We then develop a
simple model and introduce a characteristic length of defect creation which allows us to describe phenomenologically
and quantitatively the broadening of the specific heat transition. This model is then applied to the resistivity
results.

The process of defect creation in a sample irradiated by high
energy electrons is based on elastic collisions between the
incident electrons and the atomic nuclei of the crystal
\cite{vajda-77}. There is an activation energy - the threshold
energy $E_{t}$ - necessary to create stable vacancy-interstitial
defects (Frenkel pairs), i.e. vacancy and interstitial separated
enough to avoid spontaneous recombination. In the case of
monoatomic
hexagonal compact structures, $E_{t}$ is typically about a few ten eV \cite%
{vajda-77}. This value has to be compared to the maximum energy
transferred to the atom by an incident electron which is inversely
proportional to the mass of the nuclei. In UPt$_{3}$, Uranium and
Platinum are heavy nuclei with Z=238 and Z=195 respectively. The
maximum transmitted energy is respectively of 79.5 and 97.0 eV for
incident electrons of energy 2.5~MeV, and 34.1 and 41.7 eV for an
incident energy of 1.5 MeV. Therefore the fact that irradiation
with 1.5 Mev electrons has no effect on the specific heat gives us
a bottom limit for the threshold energies in UPt$_{3}$ equal
respectively to 34.1 and 41.7 eV for U and Pt atoms. These values
are quite high and might be explained by the high compactness of
the crystallographic structure in UPt$_{3}$.

The cross section for displacement of an atom by an electron with an incident energy $E$ depends both on the threshold
energy and on the energy transferred to the atoms. For thick samples, it is then necessary to take into account the
energy losses along the electron path. It is found to be equal to -2.37~keV/$\mu $m and to remain approximately
constant over a depth of 270~$\mu $m (corresponding to the thickest sample of this study). This decrease of incident
energy as a function of the penetration depth inside the sample induces a decrease of the cross section for elastic
collisions and therefore of the concentration of defects. This is a schematic mechanism from which the rate of defects
creation depends on the depth inside the sample. It is even stronger if the initial energy is barely enough for the
transmitted energy to exceed the threshold. A detailed rigorous description of rate of defect creation inside the
sample as a function of energy is certainly possible. Such analyses have been effectively performed, for example to
determine the threshold energies in the diatomic compound TaC \cite{Gosset-87}. But it is not our purpose here, which
is rather, first, to check that this is indeed responsible for the measured effects and, second, to determine which
experimental conditions have to be used in order to prevent such an inhomogeneous distribution. This is the aim of the
phenomenological model developed in the next section.\newline

\section{Phenomenological model}

In order to quantify our hypothesis, we need the concentration of defects as a function of depth in the sample ($c(x)$,
$x=0$ corresponding to the side of the sample facing directly the incident beam). As explained above, $c(x)$ which is
of course proportional to fluence, is not known exactly and depends on parameters like the energy dependence of the
cross section and the decrease of the energy of the incident electrons due to inelastic scattering. So we have chosen
to use a simple phenomenological exponential dependence, parameterised by a characteristic length for defect creation
$\lambda $ : $c(x)=c(0)exp(-x/\lambda )$.

The second element is the relationship between defects concentration and T$%
_{c}$ decrease. This is well known and studied, both experimentally and
theoretically. Indeed, it has been already demonstrated that the critical
temperature of UPt$_3$ is strongly dependent on the concentration of defects %
\cite{Vorenkamp-93,Dalichaouch-95,Kycia-98}. This feature is common to all
unconventional superconductors like high-T$_c$ oxides (see for example \cite%
{Rullier-Albenque-00}) or Sr$_2$RuO$_4$ \cite{McKenzie-80}. The decrease of
the critical temperature in a conventional or unconventional superconductor
due to a pair breaking mechanism is always described by the
Abrikosov-Gor'kov formula \cite{AG}:

\begin{equation}  \label{AG}
ln\left(\frac{T_{c}}{T_{c0}}\right)=\psi\left(\frac{1}{2}\right)-\psi\left(%
\frac{1}{2}+\frac{\alpha}{2\pi k_{B}T_{c}}\right)
\end{equation}

$T_c$ is the critical temperature, $T_{c0}$ the critical temperature of the
sample without pair breaking, $\psi$ the digamma function and $\alpha$
characterises the strength of the pair breaking mechanism. In the limit of
small pair breaking ($\alpha \rightarrow 0$), the decrease of the critical
temperature is linear and equal to $\frac{\pi\alpha}{4k_{B}}$. Larkin has
shown that for an unconventional superconductor with isotropic impurity
scattering, $\alpha$ is equal to $\frac{\hbar\Gamma}{2}$ where $\Gamma$ is
the scattering rate \cite{Larkin}. At fixed incident energy $E$, for an
infinitely thin sample, the scattering rate is proportional to the number of
defects and therefore to the fluence. As the cross section of individual
scattering centers is not known, we rather define a new parameter : $\left|%
\frac{dT_{c}}{d\Phi}(E)\right|$, the initial rate of decrease of $T_c$ with
fluence, for an infinitely thin sample. For a given material, this parameter
will only depend on the (fixed) incident energy. Then, the local critical
temperature $T_{c}(x)$ at depth $x$ in the sample is given by equation (\ref%
{AG}) with :

\begin{equation}  \label{alpha}
\alpha = \alpha(x) = -\frac{4k_{B}}{\pi}\left|\frac{dT_{c}}{d\Phi}%
(E)\right|\Phi exp(-x/\lambda)
\end{equation}

A strong test of the model is to reproduce the specific heat behaviour.
Again, exact calculations of the specific heat for a given homogeneous
concentration of defects are possible \cite{Yang-01}. They depend on the
symmetry of the order parameter, but they do not account for the double
transition. So we used a much simpler approach, which appears more
proportionate to the ambition of our model. We first make a simple fit of
the specific heat of the virgin sample, which defines a function C$_{T_{c0}}$%
(T) which respects the entropy conservation. For a homogeneous sample of
reduced temperature T$_{c}$ (due to an \emph{homogeneous} distribution of
pair breaking centers), the specific heat will be C$_{T_c}$(T). It has a
residual linear term, due to impurity bound states \cite{Scmitt-86}. This
term is estimated crudely as $\gamma_{res} = \left. \frac{C(T)}{T}%
\right|_{T\rightarrow0}=\gamma\frac{T_{c0}-T_{c}}{T_{c0}}$, where $\gamma$
is the Sommerfeld coefficient in the normal state. We then simply rescale
with respect to T$_{c}$ the specific heat of the virgin sample, using the
following expression which respects entropy balance :

\begin{equation}  \label{Cp}
\frac{C_{T_c}(T)}{T}=\gamma_{res}+\frac{\gamma -\gamma_{res}}{\gamma}\frac{%
C_{T_{c0}}(T\frac{T_{c0}}{T_c})}{T\frac{T_{c0}}{T_c}}
\end{equation}

A simple integration over the sample thickness using equations \ref{Cp}, \ref%
{alpha} and \ref{AG} gives the fit of the specific heat of irradiated
samples for each fluence, which depends only of the two free parameters $%
\lambda$ and $\left|\frac{dT_{c}}{d\Phi}(E)\right|$. On figure \ref{figCp},
the full lines are the results of the fit for $\lambda = 90\mu$m and $\left|%
\frac{dT_{c}}{d\Phi}(2.5MeV)\right|$ = 115mK/10$^{18}$ e$^-$.cm$^{-2}$. It is gratifying to see that it reproduces well
both the broadening and the small shift of the onset critical temperature for both fluences and with the same
parameters. More insight into the physical effects at work is gained looking at the distribution of $T_c$ inside the
sample deduced from these parameters.

\begin{figure}[t]
\includegraphics[bb = 48 252 477 589, width = 0.6\linewidth,
clip]{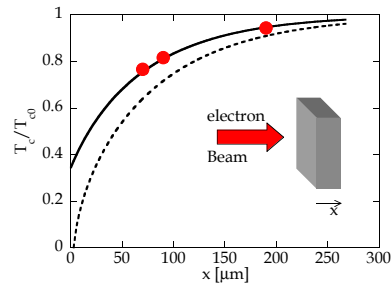} 
\caption{ Dependence of the critical temperature ($T_c$)
normalized to the critical temperature before irradiation
($T_{c0}$)with the thickness of a sample irradiated with electron
a fluence $\Phi=3.5~10^{18}$ e$^-$/cm$^2$. The closed circle are
the experimental data measured by resistivity. The solid line is
the value expected in function of the depth for an exponential
dependence of the rate of defects with the parameters (see text) $\protect%
\lambda=80~\protect\mu$m and
d$T_c$/d$\Phi$=-90~mK/$10^{18}$e$^-$/cm$^{2}$. The dashed line is
with the parameters $\protect\lambda=90~\protect\mu$m and
d$T_c$/d$\Phi$=-115~mK/$10^{18}$e$^-$/cm$^{2}$ used for the
specific heat fit on figure 2. The inset is a schematic
description of the geometry of the irradiation of the samples. }
\label{Tcvsx}
\end{figure}

Let us first discuss the T$_{c}$ change as measured by resistivity. As soon
as a thin layer of the sample is superconducting, it will short circuit the
sample. In the present geometry of the irradiation, the part of the sample
with the largest critical temperature (producing the drop to the zero
resistivity on figure \ref{figRes}) is the side opposite to the electron
beam. It should be the same part of the sample that gives the onset T$_{c}$
for the specific heat. On figure \ref{Tcvsx}, the ratio between the critical
temperature after and before irradiation with the same fluence for samples
of different thickness is reported. The solid curve is T$_{c}(x)$, where x
is the sample thickness, as deduced from equations \ref{AG} and \ref{alpha}.
The parameters differ slightly than from those of the specific heat fit ($%
\lambda = 80\mu$m and $\left|\frac{dT_{c}}{d\Phi}(2.5MeV)\right|$ = 90mK/10$%
^{18}$ e$^-$.cm$^{-2}$). For comparison, the dashed line shows the expected T%
$_c$ variation for the resistivity experiment using the parameters of the
fit of $C_{p}$. However, according to the crudeness of our model, and the
difficulties always met in the comparison of T$_{c}$ between specific heat
and resistivity, we believe that it is a "good enough" quantitative
agreement. Of course, the lines in figure \ref{Tcvsx} also represent the
distribution of T$_{c}$ as predicted by our model in a single sample. For
the intermediate fluence used in the resistivity experiment, it can be seen
that the distribution of T$_{c}$ is very large and extends very quickly from
almost T$_{c0}$ (small change in the T$_{c}$ onset for a sample 3 times
thicker than $\lambda$) to 0. The amazing broadening of the specific heat
transition is therefore simply explained by a trivial effect of sample
geometry and small $\lambda$, due to "unlucky" experimental conditions
(2.5MeV being too close to the threshold for defect creation).\newline


\section{Consequences on the transport measurements}

Let us summarize some further consequences on transport measurements. The
fact that the Sommerfeld coefficient $\gamma$ of the normal state specific
heat does not depend on the irradiation shows that the density of states at
the Fermi level is not affected. This effect seems to contradict the
increase of the quadratic term of the temperature dependence of the
resistivity ($AT^2$) under irradiation (see section I). Indeed A and $\gamma$
are related by the well known Kadowaki-Woods ratio \cite{Kadowaki-86}. One
can then speculate that the apparent increase of A is an artifact of the
distribution of defects. Indeed, when the whole sample is in the normal
state, the measured conductivity is an average of the conductivity of the
different layers of the sample, which are all in parallel owing to the
contact geometry. In the simplest models which assume that the Matthiessen's
rule is verified, only the residual resistivity should dependent on the
impurity concentration. Then, the temperature dependence of the conductivity
should be described by :

\begin{equation}  \label{resistivity apparent1}
\widetilde{\sigma}(T)=\frac{1}{t}\int_0^t\frac{dx}{\rho_0(x)+AT^2}
\end{equation}

where $t$ is the thickness of the sample, $\rho_0(x)$ the residual
resistivity at a depth $x$, and $AT^2$ the intrinsic quadratic term,
independent of defects concentration. In general the calculation has to be
done numerically, however in the limit where $AT^2$ is small compared to $%
\rho_0(x)$, the temperature dependence of the resistivity deduced from
equation \ref{resistivity apparent1} will be of the form $1/\widetilde{%
\sigma_0}+\widetilde{A}T^2$ with :

\begin{equation}
\widetilde{\sigma_0}=\langle \sigma_0 \rangle
\end{equation}
\begin{equation}
\widetilde{A}=A\frac{\langle \sigma_0^2 \rangle}{\langle \sigma_0 \rangle^2}
\end{equation}

where $\langle ...\rangle$ is the average over the depth. An increase of the
\emph{dispersion} of the defects concentration can therefore be at the
origin of an apparent increase of the quadratic temperature dependence.
Qualitatively, such an effect is observed with the increase of the thickness
of the sample, but also for an increase of the fluence for the same sample.

\begin{table}[b]
\begin{tabular}{|c|c|c|}
\hline Thickness & $\frac{\rho_0}{\rho_0^{Pure}}$ &
$\frac{A}{A^{Pure}}$ \\ ($\mu$m) & meas. (Calc.) & meas. (Calc.)
\\ \hline 70 & 9.3 (12) & 1.12 (1.03) \\ 90 & 8.7 (10.6) & 1.15
(1.05) \\ 190 & 6.9 (5.6) & 1.30 (1.12) \\ \hline
\end{tabular}%
\caption{Ratio after and before irradiation by electrons of energy
of 2.5 MeV with a fluence of 3.5 10$^{18}$ e$^-$/cm$^2$ of (a) the
residual resistivity extrapolated to T=0K and (b) the quadratic
temperature dependence of the resistivity. The values measured are
compared with the values numerically calculated (between bracket)
with an exponential dependence of the concentration of defects in
the thickness of the samples. We have also assumed that only the
residual resistivity changes (for an
homogeneous distribution of defects, see text).The parameters used are $%
\protect\lambda$=80$\protect\mu$m (see text) and a variation of
the resistivity 1.4 (0.6) $\protect\mu\Omega$.cm/10$^{18}$
e$^-$/cm$^2$ in the a (c) direction. These parameters are the same
as those used in the figure 4.}
\end{table}

Quantitatively, since the dependence of the residual resistivity for an
infinitely thin sample is proportional to the density of defects, again, a
unique parameter $\frac{\partial\rho_{0,i}}{\partial\Phi}$ (where i
represents the two different crystallographic orientations) controls the
increase of the apparent residual resistivity and of the quadratic term of a
thick sample in each direction. At the temperature T, the resistivity at
depth x inside the sample irradiated with a fluence $\Phi$ is given by:

\begin{equation}
\rho_0(x)=\rho_0+\frac{\partial\rho_{0,i}}{\partial\Phi}\Phi
exp(-x/\lambda)+AT^2
\end{equation}

where $\rho_0$ is the residual resistivity before irradiation and $\lambda$
the characteristic length already defined above. The result of this
calculation with a parameter $\frac{\partial\rho_{0,i}}{\partial\Phi}$ of
1.4 (resp. 0.6) $\mu\Omega$.cm/10$^{18}$e$^-$/cm$^2$ for a current along the
a-axis (resp. c-axis) is shown on figure 4 for the same sample after
different rate of irradiation and on the table 1 for a same fluence on
samples of different thickness. The general trend is well described.\newline

\begin{figure}[t]
\includegraphics[bb = 60 261 496 773, width = 0.6\linewidth,
clip]{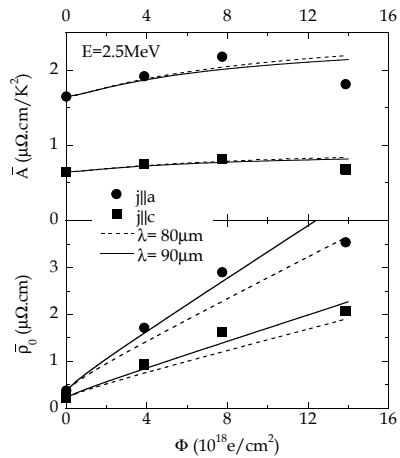} \caption{Dependence of the residual resistivity
extrapolated to T=0K (bottom) and of the quadratic temperature
dependence of the resistivity (top) for different rates of
irradiation for the two samples R4 and R5 of respective
crystallographic orientations : current along c-axis and current
along a-axis. R4 (squares) is 220$\protect\mu$m thick, and R5
(circles) is 240$\protect\mu$m thick. Lines are the dependence
calculated for a rate of defects decreasing exponentially in the
thickness of the sample with a characteristic length of
90$\protect\mu$m (plain) or 80$\protect\mu$m (dashed). Only the
residual resistivity is assumed to be modified by the defects,
with a rate of 1.4 (resp. 0.6)
$\protect\mu\Omega$.cm/10$^{18}\times $e$^-$.cm$^{-2}$ for the
current along the a (resp. c) direction.} \label{fig4}
\end{figure}

Of course, the most interesting transport measurement is the thermal
conductivity $\kappa$ in the superconducting state. The measurements of $%
\kappa$ on irradiated samples reported earlier by our group \cite{Suderow-99} have been performed on samples typically
of several hundreds $\mu m$ thick. The irradiation parameters were an incident energy of 2.5MeV and a fluence of
3.17$\times10^{18}$e$^-$/cm$^2$ and 11$\times10^{18}$ e$^-$/cm$^2$. The main result was the observation of a dramatic
decrease of the thermal conductivity in the superconducting phase and a large increase of the zero temperature
extrapolation of $\kappa /T$ with the fluence of the irradiation. The huge distribution of T$_{c}$ induced by both
irradiations in these experimental conditions (down to T$_{c}$ = 0~K) give a simple and natural explanation to these
features. So the question of a "universal limit" in UPt$_{3}$ remains completely open, and should be settled by
measurements on very thin samples irradiated on both sides to minimize the defects inhomogeneities.  From the
parameters used in our model, we estimate that for samples 70$\mu m$ thick, irradiated on both faces with total
fluences below 1.5$\times10^{18}$e$^-$/cm$^2$, the variation of $T_c$ should remain below 20$\%$ (from $T_c$=0.55~K
before irradiation to 0.46~K after), with an inhomogeneous broadening below 10$\%$ of the variation of $T_c$ (i.e.
10~mK which is small compare to the broadening of the "pure" samples).
Let us note that for the same experiments performed on the bismuth compounds %
\cite{Behnia-99}, the samples were already much thinner ($\approx$ 20$\mu m$ %
\cite{BehniaP}), the transmitted energies twice larger (due to lighter
elements) and the threshold energy 4 times smaller \cite{Florence95},
ensuring that damage was homogenously distributed through the sample in
contrast to the situation encountered for UPt$_3$. The same "good"
conditions should also be found in the compounds composed of lighter
elements like Cerium based heavy fermion systems or borocarbide
superconductors.\newline

\section{Conclusion}

In conclusion, we have measured and analyzed the specific heat of thick
samples of UPt$_3$ damaged  by high-energy electron irradiation. A strong
smearing of the specific heat jump at the  superconducting transition and a
large increase of the residual $C/T$ term are observed. We have  shown that
a strong dependence of the concentration of defects along the sample
thickness can explain  quantitatively this feature. Such inhomogeneities
have a natural explanation with the heavy  elements of UPt$_3$. This is
further confirmed by the dependence of the critical temperature  measured by
resistivity with the thickness of the sample. Such macroscopic
inhomogeneities give  also a natural explanation to the increase of the
quadratic term of the temperature dependence of  the resistivity, and to the
apparent absence of a universal limit in the heat transport  measurements
reported previously.\newline

We gratefully acknowledge useful discussion with I. Fomin, H. Suderow, V.
Mineev and N.E. Hussey. P.R. was partly supported by a Marie Curie
Fellowship of the European Community under contract n$^{\circ}$
HPMF-CT-2000-00954.

\end{document}